\documentclass[aps,prd,reprint,superscriptaddress,amsmath,amssymb,nofootinbib,natbib,longbibliography]{revtex4-2}

\usepackage{lipsum}
\usepackage{graphicx}
\usepackage{xcolor}
\usepackage[colorlinks]{hyperref}
\usepackage[capitalize]{cleveref}
\usepackage{aas_macros}

\usepackage{bm}

\usepackage{soul}

\renewcommand{\d}[2][]{\operatorname{d}^{#1}\!{#2}}

\newcommand{\Hu}{\mathcal{H}}
\newcommand{\R}{\mathcal{R}}

\newcommand{\E}{\mathcal{E}}
\newcommand{\D}{\mathcal{D}}

\usepackage{mathtools}
\usepackage{diffcoeff}
\usepackage{amsmath}

\usepackage[T1]{fontenc}
\usepackage[utf8]{inputenc}

\begin{document}

\title{Quantum initial conditions for curved inflating universes}
\author{M.~I.~Letey}
\email[]{maryletey@fas.harvard.edu}
\altaffiliation[]{equal contribution}
\affiliation{Kavli Institute for Cosmology, Madingley Road, Cambridge, CB3 0HA, United Kingdom}
\affiliation{John A. Paulson School of Engineering and Applied Sciences, Harvard University, Cambridge, Massachusetts 02138, USA}

\author{Z.~Shumaylov}
\email[]{zs334@cam.ac.uk}
\altaffiliation{equal contribution}
\affiliation{Kavli Institute for Cosmology, Madingley Road, Cambridge, CB3 0HA, United Kingdom}
\affiliation{Department of Applied Mathematics and Theoretical Physics, University of Cambridge, Wilberforce Road, Cambridge, CB3 0WA, United Kingdom}

\author{F.~J.~Agocs}
\affiliation{Center for Computational Mathematics, Flatiron Institute, 162 Fifth Avenue, New York, New York, USA}
\affiliation{Kavli Institute for Cosmology, Madingley Road, Cambridge, CB3 0HA, United Kingdom}
\affiliation{Astrophysics Group, Cavendish Laboratory, J. J. Thomson Avenue, Cambridge, CB3 0HE, United Kingdom}

\author{W.~J.~Handley}
\affiliation{Kavli Institute for Cosmology, Madingley Road, Cambridge, CB3 0HA, United Kingdom}
\affiliation{Astrophysics Group, Cavendish Laboratory, J. J. Thomson Avenue, Cambridge, CB3 0HE, United Kingdom}

\author{M.~P.~Hobson}
\affiliation{Astrophysics Group, Cavendish Laboratory, J. J. Thomson Avenue, Cambridge, CB3 0HE, United Kingdom}

\author{A.~N.~Lasenby}
\affiliation{Kavli Institute for Cosmology, Madingley Road, Cambridge, CB3 0HA, United Kingdom}
\affiliation{Astrophysics Group, Cavendish Laboratory, J. J. Thomson Avenue, Cambridge, CB3 0HE, United Kingdom}

\begin{abstract} 
    We discuss the challenges of motivating, constructing, and quantising a canonically-normalised inflationary perturbation in spatially curved universes. We show that this has historically proved challenging due to the interaction of nonadiabaticity with spatial curvature.
    We construct a novel curvature perturbation that is canonically normalised in the sense of its equation of motion, and is unique up to a single scalar parameter. With this construction it becomes possible to set initial conditions invariant under canonical transformations, overcoming known ambiguities in the literature. This corrected quantisation has potentially observational consequences via modifications to the primordial power spectrum at large angular scales, as well as theoretical implications for quantisation procedures in curved cosmologies filled with a scalar field.
\end{abstract}

\maketitle 

\section{\label{sec:intro}Introduction}
Cosmological inflation~\cite{Starobinsky1979,Guth1981,Linde1982,Albrecht1982} provides explanatory power for observations of cosmic microwave background (CMB) anisotropies~\cite{Planck2018I,Planck2018CMB,Planck2018Parameters,Planck2018Inflation}, by yielding the quantum fluctuations that seed large scale structure today~\cite{1992PhR...215..203M}. Additionally, inflation also resolves the horizon and curvature problems, both of which can be thought of as initial conditions for the universe~\cite{tasi}. From a theoretical standpoint, it is inconsistent to assume a flat universe at the start of inflation, and instead one could consider inflation as starting in a general K$\Lambda$CDM (cold dark matter with a cosmological constant and spatial curvature) universe~\cite{White1996, Lasenby2005, Bonga2016c, Bonga2017, Forconi2021}. Such investigations are further motivated in light of the conversation in the literature regarding the statistical significance (or lack thereof) of the preference in CMB and baryon acoustic oscillation (BAO) data for present-day curvature~\cite{DiValentino2019,Handley_2021,Efstathiou2020,2021PDU....3300851V,2021ApJ...908...84V,cullenhowlet}, and it is undeniably true that any observation of present-day curvature would have profound implications for theories of inflation~\cite{Kleban2012,Hergt_2022,Tegmark2007,Ellis2020}.

The imprint of quantum perturbations on the CMB as anisotropies is described by the power spectrum for the gauge-invariant comoving curvature perturbation variable $\R$. Thus, toward the goal of computing the power spectrum for $\R$ in a curved inflationary spacetime, Ref.~\cite{Handley_2019} finds the Mukhanov-Sasaki equation of motion for $\R$ for nonzero $K$, in analogy with standard inflationary calculations, previously also derived in \cite{PhysRevD.66.084009}. Introducing curvature markedly complicates the equation of motion for $\R$, which only simplifies in two important cases: First, when $K=0$; second, when one takes the matter field to be either a scalar field without potential (i.e., a stiff fluid) or hydrodynamical matter ~\cite{Garriga_1999, 1992PhR...215..203M}. During very early evolution the inflaton can be approximated as a stiff fluid, and during slow roll as hydrodynamical matter \cite{Shumaylov_2022}, thus simplifying the equation of motion. However, this does not occur in the period between the two regimes, which we show is due to the interaction of nonadiabatic perturbations with curvature. This motivates the need for a novel inflationary perturbation variable, which we introduce.

The purpose of this paper is twofold. First, we construct a field satisfying a canonical wave equation, which in the limit $K=0$ recovers the standard Mukhanov variable, allowing one to use the existing analysis of scalar fields in spaces with a Friedmann-Lema\^{i}tre-Robertson-Walker (FLRW) metric. Second, we use the proposed construction to derive initial conditions on the curvature perturbation, which are covariant and canonically invariant.

\begin{figure*}
    \includegraphics{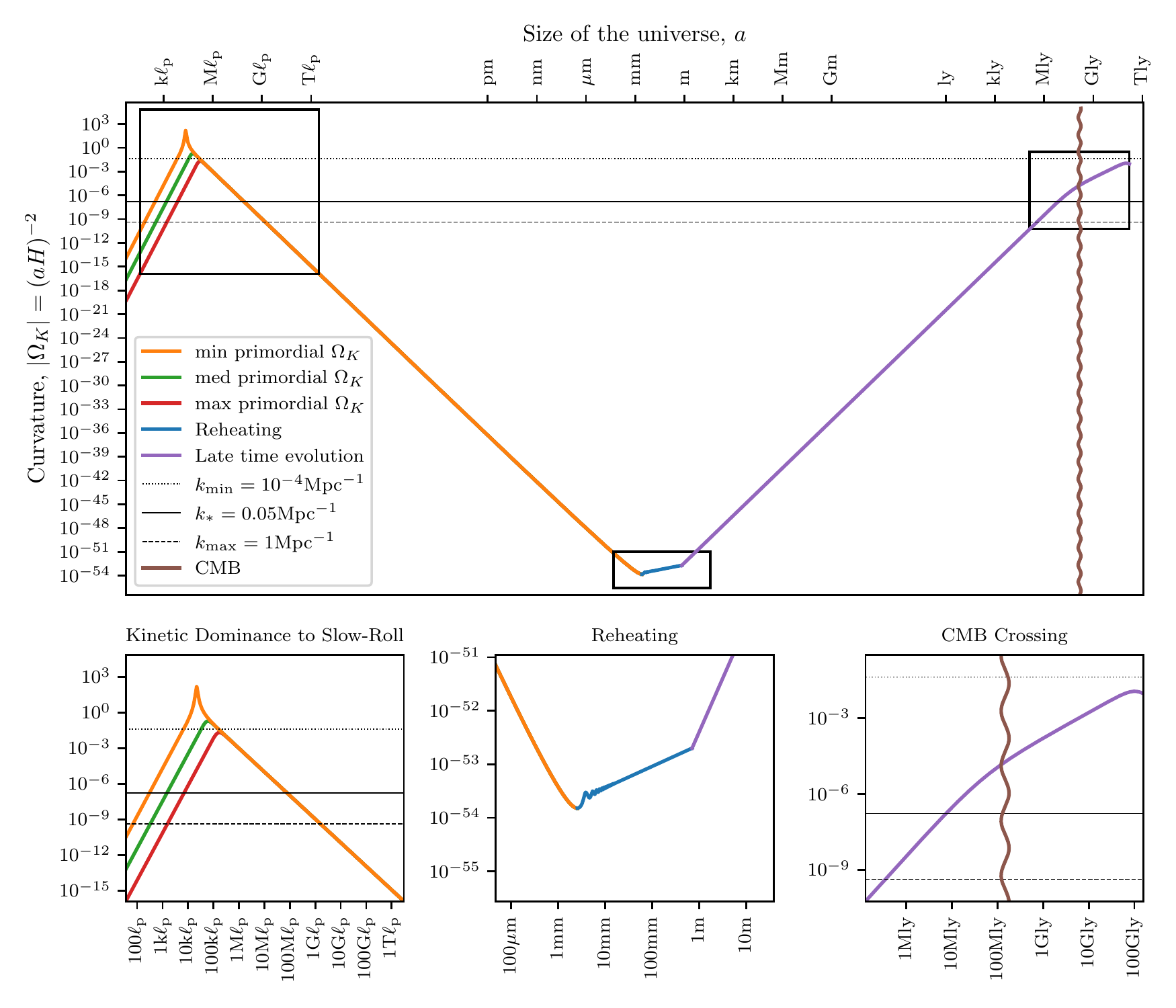}
    \caption{Timeline of a K$\Lambda$CDM cosmology using \textit{Planck} best-fit parameters~\cite{Planck2018Parameters} with \texttt{plikTTTEEE}+\texttt{lensing} for a cosmology including spatial curvature and $V(\phi)\propto\phi^{4/3}$ potential with instant reheating~\cite{Hergt_2022}. The universe begins in a kinetically dominated pre-inflationary epoch~\cite{Linde2001, Contaldi2003, Boyanovsky2006, Boyanovsky2006a, Destri2008, Ramirez2009, Destri2010, Ramirez2012, Ramirez2012a, Handley2014, Lello2014, Cicoli2014, Scacco2015, Hergt2019, Ragavendra2020}, with the comoving horizon $\frac{1}{aH}$ and curvature $\Omega_K = -\frac{K}{(aH)^2}$ parameters growing. Inflation begins when the scale factor of the universe reaches about $10^5$ Planck lengths, acting to flatten the universe and dramatically shrinking the comoving horizon through a period of slow roll. At a scale factor of about 1 mm the inflaton reaches the bottom of its potential, with oscillations about the minima simulating a matter dominated universe. At some point in between 10 mm and 10 m the universe undergoes a reheating phase which we model as instantaneous, but can easily be extended to allow greater freedom~\cite{Hergt_2022}. The universe then grows in a protracted radiation-dominated phase, undergoing several phase transitions until recombination at about the time that the universe transitions into a matter dominated epoch when the scale factor is about 0.1 Gly. Post recombination/CMB the universe eventually enters into a late time dark energy dominated epoch and the curvature and comoving horizon start to shrink again, until we reach a universe today with radius $a_0\approx 50$Gpc. The best-fit parameters today even with a small amount of curvature place strong constraints on inflationary potential consistent with this history, and only a relatively small range of primordial curvatures prove compatible. For more detail, consult \citet{Hergt_2022}.\\
    (${H_0=64.03{\text{ kms}^{-1}\text{ Mpc}^{-1}}}$, ${\Omega_m=0.3453}$, ${\Omega_K=-0.0092}$, ${\log 10^{10}A_s = 3.0336}$, ${n_s=0.9699}$, ${z_*=1089.61}$)
    \vspace{10pt}}
    \label{fig:phi4o3}
\end{figure*}

To compute the power spectrum, the curvature perturbation variable must be evolved according to its equation of motion from some initial conditions. In flat spacetimes, initial conditions from the Bunch-Davies vacuum set arbitrarily far back in the past are typically used. It is nontrivial to generalise such quantisation schemes to a curved inflationary spacetime on two counts: In the first instance, eternal inflation is impossible in the context of curvature~\cite{Hergt_2022}; refer further to \cref{fig:phi4o3} for a timeline of inflationary K$\Lambda$CDM cosmology. In the second, traditionally posed methods such as the Bunch-Davies vacuum, Danielsson vacuum, $\alpha$-vacua, and Hamiltonian diagonalisation lack canonical invariance.
Many works have previously analysed cosmological perturbations in the presence of curvature and attempted to answer the question of setting initial conditions, e.g. \cite{kiefer2022power,PhysRevD.78.043534} or in the context of bouncing inflationary models \cite{Hwang_2002, Martin_2003,lilley2011observational}, especially highlighting issues with the usual generalisation of the Mukhanov-Sasaki variable for setting initial conditions; see e.g. \cite{PhysRevD.77.083513}. However, various assumptions have to be used in order to select the ``correct'' vacuum, whether by asymptotic considerations, or by analogy with the flat case, without addressing existing issues of lack of canonical invariance of such selections \cite{VBozza_2003, giovannini2003assigning, Agocs_2020,fulling_1989,fullinggravity,grain2020canonical}. While canonical transformations preserve the (classical or quantum mechanical) evolution, they may lead to different initial states and, consequently, to different corrections in the power spectrum.

More recently a method of setting initial conditions through minimisation of the renormalised stress-energy tensor (RSET) has been proposed in \cite{Handley_2016}, which avoids making assumptions about the asymptotic behaviour of the inflationary spacetime. RSET initial conditions are similar to the instantaneous vacuum and adiabatic regularization \cite{PhysRevD.91.064051, PhysRevD.106.065015}, but unlike the latter, only require the RSET to be minimized, rather than vanish. RSET initial conditions have since been applied to inflationary collapse models \cite{BENGOCHEA2017338}, where until now, initial conditions were selected only according to their backreaction \cite{giovannini2003assigning}.  Moreover, such initial conditions have been shown to be canonically invariant in flat universes~\cite{Agocs_2020}.

The paper is structured as follows. For clarity, we start by summarizing known results in \cref{sec:background,sec:fluid}: in \cref{sec:background} we introduce standard cosmological perturbation theory generalised to curved spacetimes, set up generalised equations for $\R$, and discuss vacuum selection; \cref{sec:fluid} provides formalism clarifying what prevents $\R$ from having a canonically normalised wave equation of motion in the presence of curvature. This motivates \cref{sec:zeta}, in which we propose a novel curvature perturbation variable that admits an equation of motion of a simple harmonic oscillator. This allows for connections to established work in quantum fields in curved spacetimes, in addition to allowing this variable to be easily quantised by the minimised-RSET procedure given in \cref{sec:rst}, which is a generalisation of the procedure proposed by Ref.~\cite{Handley_2016}. These calculations are as general as possible, written in terms of inflationary sound speed, allowing for extensions to nonstandard inflationary Lagrangians. Finally, \cref{sec:powerspectrum} discusses the use of this novel variable as a means of setting initial conditions, not only for $\R$ but for any first order scalar perturbation variable. The resulting power spectrum for $\R$ is plotted, and we discuss and draw conclusions in \cref{sec:conclusion}.

\section{\label{sec:background}Background}
We begin with a discussion of the relevant inflationary perturbation theory setup, summarising previous results before motivating the need for both an alternate curvature perturbation variable and a more robust method of quantisation. 

By convention, we will work in conformal time $\eta$, where we will denote $' \equiv \text{d}/\text{d}\eta,$ and use the conformal Hubble parameter $\Hu = aH = a'/a.$

\subsection{\label{sec:pert}Perturbation theory setup}

To leading order, both spacetime and the inflaton field can be taken to be homogeneous and isotropic, i.e. dependent only on $\eta.$ We will consider the background to be a general FLRW spacetime, whose metric is given by
\begin{equation}
    \d{s}^2 = a^2(\eta)\{\d{\eta}^2 - c_{ij}\d{x}^i \d{x}^j \},
\end{equation} 
with the spatial metric 
\begin{equation}
    c_{ij}\d{x^i}\d{x^j} = \frac{\d{r}^2}{1-Kr^2} + r^2(\d{\theta}^2  + \sin^2\theta\d{\phi}^2).
    \label{eq:kfrw}
\end{equation} 
This background is then subject to scalar perturbations 
\begin{gather}
    \d{s}^2 = a^2(\eta) \{ (1+2\Phi)\d{\eta}^2 
    - \left( (1-2\Psi)c_{ij}\right)\d{x}^i \d{x}^j \},
    \label{eq:fullpert}
\end{gather} 
where we have picked the Newtonian gauge and omitted vector and tensor perturbations as they decouple from the scalar perturbations. See Ref.~\cite{tasi} for a more in-depth discussion of gauge choices. We restrict our attention to scalars in this instance since vectors generically decay during inflation, and tensors are already canonically normalised, even for $K\neq 0$~\cite{Handley_2019}.

The inflaton is taken to be a scalar field minimally coupled to the curved spacetime,
\begin{equation}
    S = \int \d[4]{x}\sqrt{|g|}\left\{ \frac{1}{2}R + \frac{1}{2}\nabla^\mu\phi\nabla_\mu\phi - V(\phi)\right\}.
    \label{eqn:inflatonaction}
\end{equation}
which is homogeneous and isotropic to zeroth order and is perturbed by $\delta \phi (\eta,\text{\textbf{x}})$. The usual background inflation dynamics are given by
\begin{align}
    \Hu^2 &= \frac{a^2}{3}\rho - K,
    \label{eq:hubblesquare} \\
    \Hu' &= \Hu^2 + K - \frac{a^2}{2}(\rho + p),
    \label{eq:hubbleprime}
\end{align}
where \eqref{eq:hubblesquare} and \eqref{eq:hubbleprime} are the Friedmann and Raychaudhuri equations respectively. The background energy density $\rho$ and pressure $p$ for the inflaton are
\begin{align}
    \rho &= \frac{1}{2a^2}(\phi ')^2 + V(\phi),
    \label{eq:rho}\\
    p &= \frac{1}{2a^2}(\phi ')^2 - V(\phi).
    \label{eq:p}
\end{align} 

\subsection{\label{sec:R}Gauge-invariant curvature perturbation}

Typical analyses now proceed by defining the gauge-invariant comoving curvature perturbation, 
\begin{equation}
    \mathcal{R} = \Psi +\frac{\Hu}{\phi'}\delta\phi.
    \label{eqn:Rdef}
\end{equation} 
$\R$ is of interest as it controls the perturbations to spatial curvature, as seen from the spatial Ricci scalar
\begin{equation}
    R^{(3)} = 6\frac{K}{a^2} + \frac{4}{a^2}(\nabla^2 + 3K)\R.
\end{equation}
These perturbations are directly related to CMB anisotropies, as variations in curvature at the time of last scattering will correspond to the scattered photons being more or less redshifted, resulting in the CMB temperature distribution~\cite{Lesgourgues_2014,tasi}. 

As derived in standard references~\cite{tasi}, the resulting equation of motion for $\R$ when $K=0$ is
\begin{gather}
    (z\R)'' - \left(\nabla^2 + \frac{z''}{z}\right)(z\R) = 0 \label{eq:k0r}\\
    z = \frac{a \phi'}{\Hu}.
\end{gather}
Note that here $\nabla^2 \equiv c^{ij}\nabla_i\nabla_j$ refers to the Laplacian defined by the curved spatial metric $c_{ij}.$ 

However for $K \neq 0$, by expanding the Einstein and conservation equations to first order in the perturbation variables, Ref.~\cite{Handley_2019} shows that $\R$ obeys the second-order Mukhanov-Sasaki equation
\begin{gather}
    \begin{aligned}
        0=&(\mathcal{D}^2 - K\mathcal{E})\R '' + 2\left( \frac{z'}{z}\mathcal{D}^2 - K \Hu \mathcal{E} \right)\R ' \\
        +&\left( K\left(1+\mathcal{E} - \frac{2}{\Hu}\frac{z'}{z}\right)\mathcal{D}^2 + K^2\mathcal{E}-\mathcal{D}^4 \right)\mathcal{R},
        \label{eqn:mukhanov_sasaki}
    \end{aligned}\\
    \mathcal{D}^2 = \nabla^2 + 3 K, \qquad
    \mathcal{E} = \frac{a^2(\rho+p)}{2\Hu^2}.
    \label{eqn:definitions}
\end{gather} 
This can be written more concisely by first mode-decomposing into Fourier space, namely writing
\begin{equation}
    \nabla^2 Y_k(\text{\textbf{x}}) = -\kappa^2(k)Y_k(\text{\textbf{x}}) 
\end{equation}
for $Y_k(\text{\textbf{x}})$ the appropriate hyperspherical Bessel functions~\cite{Lesgourgues_2014,birrell_davies_1982} that give the eigenspectrum of the spatially curved Laplacian. Thus we identify $\mathcal{D}^2$ with $-\kappa^2(k)+3K$, where
\begin{align}
    \kappa^2(k) &= k^2,    & k \in \mathbb{R}, k > 0:& & K&=0,-1, \nonumber\\
    \kappa^2(k) &= k(k+2), & k\in\mathbb{Z}, k > 2:&   & K&=+1,\label{eqn:kappadef}
\end{align}
Define wavenumber-dependent $\mathcal{Z}$ as
\begin{equation}
    \mathcal{Z} = z\sqrt{\frac{\mathcal{D}^2}{\mathcal{D}^2-K\mathcal{E}}} = a\sqrt{\frac{2\mathcal{D}^2\mathcal{E}}{\mathcal{D}^2-K\mathcal{E}}}
\end{equation}
\Cref{eqn:mukhanov_sasaki} can be recast as
\begin{equation}
    (\mathcal{Z}\R_k)'' + \left(\kappa^2 - \frac{\mathcal{Z}''}{\mathcal{Z}} - 2K - \frac{2K\mathcal{Z}'}{\mathcal{H}\mathcal{Z}} \right)(\mathcal{Z}\R_k) = 0,
    \label{eqn:mukhanov_sasakiwavespace}
\end{equation}
where $\R_k(\eta)$ is the Fourier component of $\R(\eta,\text{\textbf{x}})$ with wave number of magnitude $k$. 

To connect this theoretical framework to observation, we must compute the primordial power spectrum from $\R$ at horizon crossing. This is an initial value problem: $\R$ can be numerically computed according to its equations of motion \cref{eqn:mukhanov_sasaki,eqn:mukhanov_sasakiwavespace} from some initial conditions $\R(\eta_0), \R'(\eta_0)$, or equivalently, a vacuum state for the corresponding quantised variable. The correct theoretical choice for initial conditions in this case is far from clear~\cite{tasi,fulling_1989,fullinggravity}; we will now discuss possible vacuum choices.

\subsection{\label{sec:vacuum}Vacuum choice}

For quantum fields on a curved spacetime, the choice of vacuum state is either physically unclear or ambiguous. At zeroth order, the slow-roll inflaton mimics a positive cosmological constant, and thus the background dynamics of an inflationary FLRW spacetime are well-described by de Sitter space. This is true for general $K$. Since de Sitter space is maximally symmetric, there exists a natural choice of vacuum for a scalar field in this spacetime by means of the Bunch-Davies (BD) vacuum~\cite{1978field}, which is invariant under all isometries of de Sitter space. This motivates the BD vacuum as a physically reasonable choice. An alternative choice is setting initial conditions by means of Hamiltonian Diagonalisation (HD). However, it is questionable whether this choice of vacuum is physically meaningful in an expanding spacetime as the time-dependent Hamiltonian will yield infinite particle density after the instant the vacuum is set~\cite{fulling_1989,fullinggravity}. 

In the $K=0$ case, the BD vacuum can be painlessly applied to quantise the Mukhanov variable $z\R$, as its equation of motion \cref{eq:k0r} is analogous to the resulting Klein-Gordon equation of motion for a scalar field, with a time-dependent mass term given by $z$. However, applying the BD vacuum to quantise perturbations in inflation requires the spacetime to be quasi--de Sitter at the time of quantisation, which is not possible for theories of finite inflation~\cite{Handley2014,Hergt2019,Hergt_2022}. This can be seen further in \cref{fig:phi4o3}, which illustrates a period of kinetic dominance of the inflaton. 

Further, in the curved case, $\R$ no longer possesses a canonically-normalised action nor equation of motion, i.e., it does not behave like a simple harmonic oscillator (SHO) as \cref{eq:k0r}. This can be seen from the $k$-dependence of $\mathcal{Z}$ in \cref{eqn:mukhanov_sasakiwavespace}. As such, we can make no connections to the large body of work on quantum fields in curved spacetime in order to provide insight on what initial conditions to use. Almost all existing literature regarding second quantisation in curved spacetimes deal with a massive scalar field with constant coupling to gravity~\cite{parker_toms_2009,fulling_1989}, admitting an SHO-like equation of motion to then quantise the scalar field by analogy with the time-independent quantum harmonic oscillator. Thus, the first step in generalising inflationary theories to nonflat primordial curvature is to find a curvature perturbation variable that obeys an analogous SHO equation of motion, unlike $\R$. 

Additionally, as shown by Ref.~\cite{Agocs_2020} for $K=0$, neither the BD vacuum nor the vacuum from HD are robust against canonical transformations. Namely, under a canonical transformation of phase space preserving the field's equation of motion, these vacuum setting procedures yield ambiguous vacuum initial conditions that would be observationally distinguishable. Another potential vacuum choice is the one proposed by Danielsson~\cite{danielsson}. This vacuum is derived in the Heisenberg picture for the field operators, and initial conditions are set by considering the time at which each mode reaches Planckian lengths. Unfortunately, this choice is also not invariant under canonical transformations when $K = 0$.

However, as proposed in Ref.~\cite{Handley_2016}, one can instead set the vacuum by minimising the RSET. This formulation avoids consideration of the tricky concept of particles. Furthermore, it does not require any assumptions about the asymptotic behaviour of the inflationary spacetime, and so allows for noneternal theories of inflation. More crucially, this method yields canonically invariant vacuum conditions~\cite{Agocs_2020} when $K=0$. Based on calculations in \cref{sec:rst} and \cref{sec:rstcalc}, this persists even for the case of $K\neq0$, which motivates our calculations of initial conditions  resulting from this procedure. 

As discussed by \citet{fulling_1989}, computing the correct form of the RSET for a given field subject to a general action is challenging, and in the case of $\R$, is virtually intractable due to the convoluted form of its action~\cite{Handley_2019},
\begin{align}
    S^{(2)}_{\R} = \frac{1}{2}\int& \d{\eta}\d[3]{x}a\sqrt{|c|} \frac{(\phi')^2}{\Hu^2}\Bigg\{\R\mathcal{D}^2\R
     \nonumber\\
 &+ \left(\R' - K\frac{\R}{\Hu}\right)\frac{\mathcal{D}^2}{\mathcal{D}^2 -K\mathcal{E}}\left(\R' -K\frac{\R}{\Hu}\right)\Bigg\}. \label{eq:actionR}
\end{align}
As discussed further in \cref{sec:rst}, a more feasible task is finding the RSET for a massless, minimally coupled scalar field $\psi$ on a curved spacetime, with resulting equation of motion
\begin{equation}
    (a\psi)'' - \left(\nabla^2 + \frac{a''}{a}\right)(a\psi) = 0 \label{eq:eomscalarfield}
\end{equation}
for which the renormalised stress-energy tensor has been derived by \citet{birrell_davies_1982}. As proposed and discussed in Ref.~\cite{Handley_2016}, when $K=0$, an analogy can be drawn between \cref{eq:eomscalarfield} and the equation of motion \cref{eq:k0r} for $\R$ in flat space, by noting that during flat slow-roll inflation 
\begin{equation}
    \frac{a''}{a} \approx \frac{z''}{z}.
\end{equation}
Thus $a\psi$ and $z\R$ share an equation of motion, and so $\R$ can be quantised directly through the minimised RSET conditions for this arbitrary scalar field $\psi.$ 

However, for $K \neq 0$, $\R$ cannot yet be quantised using the minimised-RSET proceedure as above, since it does not have the equation of motion of a SHO. Thus, in what follows, we will motivate and derive a novel perturbation variable that obeys a canonically normalised wave equation of motion. Finally, we will derive initial conditions for this variable in \cref{sec:rst} by means of the minimised-RSET procedure in Ref.~\cite{Handley_2016} which we have generalised to curved spacetimes. 

\section{\label{sec:fluid}Motivation for a New Variable} 
As discussed above, we aim to construct a scalar perturbation variable obeying a SHO equation of motion analogous to \cref{eq:k0r} in order to make connections with the existing literature concerning quantum fields in curved spacetime and their quantisation, and further, to be able to apply minimised-RSET as a well-motivated choice of vacuum selection. 

Progress toward finding a canonically normalised perturbation variable has been made by \citet{brechetfluid}, in which the proposed variable obeys an equation of motion that recovers the wave equation seen for $\R$ in the $K=0$ case. 

To understand the results of this paper as they relate to curved inflation, let us set up again a perturbed spacetime, filled instead by a perfect fluid. Note that, with care, a scalar field can be seen as a special case of a perfect fluid to zeroth order~\cite{Faraoni_2012}. The stress-energy tensor for this component fluid can be expanded to first order as
\begin{align}
    T_0^0 &= -\rho - \delta\rho
    \nonumber \\
    T_i^j &= -(p+\delta p)\delta^j_i
    \label{eq:fluidT} 
\end{align} 

Again, we consider a perturbed spacetime in the Newtonian gauge,
\begin{equation}
ds^2=(1+2\Phi)dt^2-a^2(1-2\Psi)c_{ij}dx^idx^j
\label{eq:metricfluid}
\end{equation}

Following \citet{Unnikrishnan2010}, by expanding the Einstein field and conservation equations to first order in perturbation variables $\delta\rho, \delta p, \Phi, \Psi,$ given component tensor \cref{eq:fluidT} and metric \cref{eq:metricfluid}, we find the Bardeen equation of motion for Bardeen potential $\Phi$ to be
\begin{align}
&\Phi''+3\Hu\left(1+c_a^2\right)\Phi' - c_a^2\nabla^2\Phi
\nonumber \\
&+ \left[(1+3c_a^2)\Hu^2-K(1+3c_a^2)+2\Hu'\right]\Phi = 0,
\label{eq:bardeenperfect}
\end{align}
where
\begin{equation}
    c_a^2 \equiv \frac{p'}{\rho'} 
\end{equation}
is the adiabatic sound speed.

Adapted to the notation used above, \citet{brechetfluid} then defines an alternate comoving curvature perturbation scalar
\begin{align}
    \zeta_{\text{PF}} &= \R - \frac{2K}{a^2(\rho+p)}\Phi.
    \label{eq:zeta0}
\end{align}
Note briefly that $\zeta_{\text{PF}} = \R$ when $K=0$, and that $c_a^2 \approx -1$ during slow-roll inflation. 

The equation of motion for $\zeta_{\text{PF}}$ can then be found simply by writing $\zeta_{\text{PF}}$ in terms of $\Phi$ as
\begin{equation}
    \zeta_{\text{PF}} = \Phi +\frac{2\Hu}{a^2(\rho+p)}(\Hu\Phi+\Phi')-\frac{2K}{a^2(\rho+p)}\Phi.
\end{equation}
\Cref{eq:bardeenperfect} then reduces to the wave equation
\begin{gather}
    (z_0\zeta_{\text{PF}})'' - \left(c_a^2\nabla^2 v + \frac{z_0''}{z_0}\right)(z_0\zeta_{\text{PF}}) = 0
    \label{eq:vperfectfluid}\\
    z_0 = \frac{a^2\sqrt{\rho + p}}{c_a \Hu}.
    \label{eq:z} 
\end{gather}
Thus, given the above canonically normalised wave equation of motion for $\zeta_{\text{PF}}$, we can canonically quantise curvature perturbations for a perfect fluid on a curved spacetime, and set initial conditions using minimised RSET as discussed in \cref{sec:rst}. 

This framework can then be applied  to the specific case of the inflaton, by calculating $\rho_{\phi}, p_{\phi}, \delta\rho_{\phi}, \delta p_{\phi}, \delta\Sigma_{\phi}$ from the stress-energy tensor for a perturbed \emph{inflationary} scalar field $\phi(\eta) + \delta\phi(\eta,\text{\textbf{x}})$. 

For the sake of generality, we consider the following extension of action \cref{eqn:inflatonaction}:
\begin{equation}
    S=\int d^4x\sqrt{-g}\left[\frac12 R - P(X,\phi)\right],
    \label{eq:pofxphi}
\end{equation}
where 
\begin{equation}
    X = \frac12 g^{\mu\nu}\nabla_{\mu}\phi\nabla_{\nu}\phi. \\
\end{equation}

A more general treatment of inflationary $P(X,\phi)$ theories can be seen in Refs.~\cite{Shumaylov_2022,Garriga_1999}. Applying the results in this paper to alternate-Lagrangian theories of inflation, such as Dirac-Born-Infeld (DBI) inflation, could be a fruitful future extension of this work. 

Then, following \citet{Garriga_1999}, we define the inflationary sound speed, or the effective sound speed of the perturbations,
\begin{equation}
    c_s^2 \equiv \frac{\partial_X P}{\partial_X P + 2X\partial_{XX}P}.
    \label{eq:cs}
\end{equation}
Note that the inflaton is typically described by a standard Lagrangian 
\begin{equation}
    P(X,\phi) = X - V(\phi).
    \label{eq:usualP}
\end{equation}
Thus, usually $c_s^2 = 1$. 

\Cref{eq:cs} allows us to recast the equivalent nonadiabatic pressure perturbation for a scalar field as
\begin{equation}
    \delta p_{\text{en}} = \frac{2}{a^2} (c_s^2-c_a^2) \mathcal{D}^2\Phi
    \label{eq:entropicpphi}
\end{equation} 

\Cref{eq:cs,eq:entropicpphi} allow us to write the Bardeen equation of motion for a scalar field as
\begin{align}
&\Phi''+3\Hu\left(1+c_a^2\right)\Phi' - c_a^2\nabla^2\Phi
\nonumber \\
&+ \left[(1+3c_a^2)\Hu^2-K(1+3c_a^2)+2\Hu'\right]\Phi = \frac{a^2}{2}\delta p_{\text{en}}
\label{eq:bardeenscalar-pen}
\end{align} 
We highlight that this only differs from \cref{eq:bardeenperfect}, the Bardeen equation of motion for a spacetime with component perfect fluid,  by the addition of the entropic pressure term $\frac{a^2}{2}\delta p_{\text{en}}$ on the right-hand side. 

Given $\Phi$ described by \cref{eq:bardeenperfect}, i.e., for the toy universe filled with a perfect fluid, we have
\begin{equation}
    \zeta_{\text{PF}}' = \frac{2\Hu c_a^2}{a^2 (\rho+p)} \nabla^2 \Phi.
    \label{eq:zeta0primeperfect}
\end{equation}
However, for a perturbed FLRW universe filled instead with the inflationary scalar field, where $\Phi$ is described by \cref{eq:bardeenscalar-pen}, 
\begin{equation}
    \zeta_{\text{PF}}' = \frac{2\Hu c_s^2}{a^2 (\rho+p)} \nabla^2 \Phi + 6\frac{K\Hu(c_s^2-c_a^2)}{a^2 (\rho + p)}\Phi,
    \label{eq:zeta0primescalar}
\end{equation} 

The discrepancy between \cref{eq:zeta0primeperfect} and \cref{eq:zeta0primescalar} suggests that a SHO wave equation of motion is impossible for this variable unless $K = 0$ or $c_a^2 = c_s^2$. Writing \cref{eq:zeta0primescalar} more explicitly as 
\begin{equation}
    \zeta_{\text{PF}}' = \frac{2\Hu c_a^2}{a^2 (\rho+p)} \nabla^2 \Phi + \frac{2\Hu}{\rho+p}\delta p_{\text{en}},
    \label{eq:zeta0primescalar-pen}
\end{equation} we see that the term preventing the desired equation of motion is proportional to the entropic pressure perturbation, which is dependent on spatial curvature through $\mathcal{D}$.

\section{\label{sec:zeta}Novel Curvature Variable}
First, the Bardeen equation (\ref{eq:bardeenscalar-pen}), which fully describes the evolution of perturbations in a curved FLRW spacetime with a perturbed inflaton, is equivalent to
\begin{align}
&\Phi''+3\Hu\left(1+c_a^2\right)\Phi' - c_s^2\nabla^2\Phi
\nonumber \\
&+ \left[(1+3c_a^2)\Hu^2-K(1+3c_s^2)+2\Hu'\right]\Phi = 0.
\label{eq:bardeenscalar2cs}
\end{align}

We aim to define a variable whose dynamics simplify to a canonically normalisable wave equation under \cref{eq:bardeenscalar2cs}. Thus, consider
\begin{alignat}{2}
      \zeta &= g(\eta)\Phi&& +\frac{2\Hu g(\eta)}{a^2(\rho+p)}(\Hu\Phi+\Phi')-\frac{2Kf(\eta)}{a^2(\rho+p)}\Phi,
      \label{eq:zetadef}
\end{alignat}
where we recall
\begin{alignat}{2}
    \R &= \Phi && +\frac{2\Hu}{a^2(\rho+p)}(\Hu\Phi+\Phi'),\\
    \zeta_{\text{PF}} &= \Phi&& +\frac{2\Hu}{a^2(\rho+p)}(\Hu\Phi+\Phi')-\frac{2K}{a^2(\rho+p)}\Phi.
\end{alignat} 

So far, $g(\eta)$ and $f(\eta)$ are yet unspecified functions, and so $\zeta$ can represent any linear combination of $\Phi,\Phi'$, i.e., of the form $A(\eta)\Phi+B(\eta)\Phi'$. 
Note that currently this is the most general form for $\zeta$, since it must be a first order perturbation, and thus a linear combination of other gauge-invariant quantities. 

We can see that under \cref{eq:bardeenscalar2cs}, the derivative of $\zeta$ simplifies as 
\begin{align}
   \zeta' = &\frac{2\Hu g}{a^2 (\rho+p)} \Bigg[c_s^2\D^2 \Phi +\Phi'\Big(\frac{K}{\Hu}-\frac{f}{g}\frac{K}{\Hu}+\frac{g'}{g}\Big)+\nonumber\\
   \Phi\Bigg(&K+\frac{g'}{g}\left(\Hu+\frac{a^2(\rho+p)}{2\Hu}\right)+\nonumber\\&\qquad\frac{K}{\Hu}\frac{f}{g}\left(2\Hu+\frac{(\rho+p)'}{(\rho+p)}-\frac{f'}{f}\right)\Bigg)\Bigg]. \label{eq:fullzetaderivative}
\end{align} 

Now, by inspection of \cref{eq:fullzetaderivative} and by analogy with \cref{eq:zeta0primeperfect}, we will want an ansatz where $\zeta'$ is proportional to only $\D^2\Phi$, and not to $\Phi$ or $\Phi'$. 
\begin{equation}
    \zeta' = g\frac{2\Hu c_s^2}{a^2 (\rho+p)} \mathcal{D}^2 \Phi.
    \label{eq:zetaprime}
\end{equation}
Once this is picked as an ansatz, we are guaranteed to arrive at a wave equation, as long as we can solve equations specifying functions $f(\eta),g(\eta)$:
\begin{gather}
    f(\eta) = g'\frac{\Hu}{K}+g \label{eq:feq}\\
    \frac{f'}{f} = K+\frac{g'}{g}\left(\Hu+\frac{a^2(\rho+p)}{2\Hu}\right)+\frac{K}{\Hu}\frac{f}{g}\left(2\Hu+\frac{(\rho+p)'}{(\rho+p)}\right). \label{eq:long-gdef}
\end{gather} 
Both of these equations can be rewritten to simplify by taking $\mathcal{G}, b$ such that
\begin{align}
    \mathcal{G} &= \frac{b'}{b},
    \label{eq:beq} \\
    \mathcal{G}^2-\mathcal{G}' &= \Hu^2-\Hu'+K,
    \label{eq:Geq}
\end{align} 
and setting 
\begin{align}
   \label{eq:gdef}
    g(\eta)&=\frac{a}{\Hu}\frac{\mathcal{G}}{b},\\
    f(\eta)&= g'\frac{\Hu}{K}+g,
    \label{eq:fdef}
\end{align}
which we will use as the definition for our functions $f(\eta)$ and $g(\eta)$. We further motivate the ansatz \cref{eq:zetaprime} in \cref{sec:variable_explained}. 

For $\zeta$ defined such that \cref{eq:zetaprime} holds, the resulting equation of motion simplifies to 
\begin{gather}
    (z_g\zeta)''+\left(c_s^2\mathcal{D}^2-\frac{z_g''}{z_g}\right)(z_g\zeta)=0,
    \label{eq:finallywaveeq}\\
    z_g(\eta) = \frac{z}{g} = \frac{a^2(\rho+p)^\frac12}{gc_s\Hu},
\end{gather}
as expected from \cref{sec:fluid}. 
The only thing left is to see whether a solution $\mathcal{G}$ satisfying \cref{eq:Geq} exists. By rewriting \cref{eq:Geq} using \cref{eq:beq}, we can see that we have a linear second order differential equation for $1/b$:
\begin{equation}
b\left(\frac{1}{b}\right)''=a\left(\frac{1}{a}\right)''+K.
\label{eq:bina}
\end{equation}
Unfortunately, \cref{eq:bina} does not have a closed form solution, unless $K=0$. We can solve it numerically, however. To justify a selection of initial conditions for this differential equation, we should note, that there is an overall scaling freedom in the definition of $b$ and therefore there is only one effective degree of freedom. Furthermore, selection of initial conditions on $b$ and $\dot{b}$ at some initial time $t_0$ is equivalent to picking initial values of $f$ and $g$ at time $t_0$ as can be seen from 
\begin{align}
    \dot{b}_0 &= \frac{g_0H_0}{a_0}b_0^2\\
    b_0 &= a_0\frac{\dot{H}_0-\frac{K}{a_0^2}}{\frac{K}{a_0^2}(f_0-g_0)+g_0\dot{H}_0}, 
\end{align}
where subscript 0 corresponds to value at initial time $t_0$. Therefore, picking which initial conditions to consider for the variable $\zeta$ is equivalent to asking which variable we want to quantise at $t_0$. Note, that the question of global existence of solutions to \cref{eq:bina} is not obvious; however for \cref{sec:powerspectrum} only the local existence is of interest for derivation of the initial conditions.

With the wave equation in place, we can discuss a few things of note about the new variable. 
First of all, $\zeta$ defined via \cref{eq:fdef,eq:gdef,eq:beq,eq:Geq,eq:zetadef} has an overall arbitrary scaling, due to a scaling freedom of both $g$ and $f$. However, the resultant Mukhanov variable $v=z_g\zeta$ does not, and furthermore always collapses to the original flat Mukhanov variable $v_{\text{flat}}=z\R$ when $K=0$, even when we pick $b\not\propto a$ as the solution to \cref{eq:bina}.

We have thus constructed $\zeta$, such that the equation of motion for $\zeta$ has canonically normalisable wave equation form, and can be quantised using the minimised-RSET vacuum conditions derived below in \cref{sec:rst}.

\section{\label{sec:rst} Vacuum Conditions via RSET} 
In what follows, we will demonstrate how to apply minimised RSET to quantise $\zeta$ with $k$-space equation of motion
\begin{equation}
    (z_g\zeta_k)'' + \left(c_s^2(\eta)\kappa_{\mathcal{D}}^2(k) - \frac{z_g''}{z_g} \right)(z_g\zeta_k) = 0 
    \label{eq:zetakspace}
\end{equation}
where $\kappa_{\mathcal{D}}$ gives the wave space decomposition of the $\mathcal{D}^2$ operator
\begin{equation}
    -\mathcal{D}^2 \leftrightarrow \kappa_{\mathcal{D}}^2(k) = \kappa^2(k) - 3K
\end{equation}
We highlight that none of the following calculations are particular to the definitions of $z_g$, $\zeta$, and $c_s$, so this procedure is easily applicable for the quantisation of a wider class of variables with comparable equation of motion. 

Compare \cref{eq:zetakspace} with a massless minimally coupled scalar field given by
\begin{equation}
    S = \frac12 \int d^4x \sqrt{|g|} \left(g^{\mu\nu}\nabla_{\mu}\psi\nabla_{\nu}\psi \right).
\end{equation}
Note that this is \textit{not} the inflaton scalar field, but is merely introduced for computational convenience; the RSET for such $\psi$ have been calculated by \citet{birrell_davies_1982}, while such a calculation for $\zeta$ is not yet obvious. $\psi$ has an equation of motion in $k$-space given by
\begin{equation}
    (a\psi_k)'' + \left(\kappa^2(k) - \frac{a''}{a}\right)(a\psi_k) = 0,
\end{equation}
i.e., another wave-equation of motion with a mass function given by the scale factor. As discussed in depth in \cref{sec:rstcalc}, and in analogy with Ref.~\cite{Agocs_2020}, we introduce four extra degrees of freedom in the form of time redefinitions and field rescalings 
\begin{align}
    \eta &\longrightarrow \eta_{\zeta}, \label{eq:time-redef-zeta}\\
    \zeta &\longrightarrow \chi_{\zeta} = \frac{\zeta(\eta_{\zeta})}{h_\zeta(\eta_{\zeta})} \label{eq:field-redef-zeta}\\
    \eta &\longrightarrow \eta_{\psi}, \label{eq:time-redef-psi} \\
    \psi &\longrightarrow \chi_{\psi} = \frac{\psi(\eta_{\psi})}{h_{\psi}(\eta_{\psi})} \label{eq:field-redef-psi}
\end{align}
We show there exist unique $h_{\zeta},h_{\psi},\eta_{\zeta},\eta_{\psi}$ such that the redefined $\chi$ fields corresponding to $\zeta$ and $\psi$ have identical wave equations of motion. With this, we can map modes of $\zeta$ onto modes of $\psi$, and thus map initial conditions of $\psi$ onto initial conditions of $\zeta$. More details can be found in \cref{sec:rstcalc}. Thus, generalising the calculations performed in Ref.~\cite{Agocs_2020}, we find initial conditions for $\zeta$ at $\eta = \eta_0$ to be
\begin{align}
    |\zeta(\eta_0)|^2 &= \frac{1}{2c_s(\eta_0)z_g^2(\eta_0)\kappa_{\mathcal{D}}}, 
    \nonumber \\
    \frac{\zeta'}{\zeta}(\eta_0) &= -i\kappa_{\mathcal{D}} + \frac{a'}{a}(\eta_0) - \frac{z_g'}{z_g}(\eta_0) - \frac12\frac{c_s'}{c_s}(\eta_0) \label{eq:zetarst-conformal}
\end{align}
or in normal time at $t = t_0$ by
\begin{align}
    |\zeta(t_0)|^2 &= \frac{1}{2c_s(t_0)z_g^2(t_0)\kappa_{\mathcal{D}}}, \nonumber \\
    \frac{\dot{\zeta}}{\zeta}(t_0) &= -i\frac{1}{a(t_0)}\kappa_{\mathcal{D}} + \frac{\dot{a}}{a}(t_0) - \frac{\dot{z_g}}{z_g}(t_0) - \frac12\frac{\dot{c_s}}{c_s}(t_0)
    \label{eq:generalrsttime}
\end{align}

\begin{figure*}
    \centering
    \includegraphics{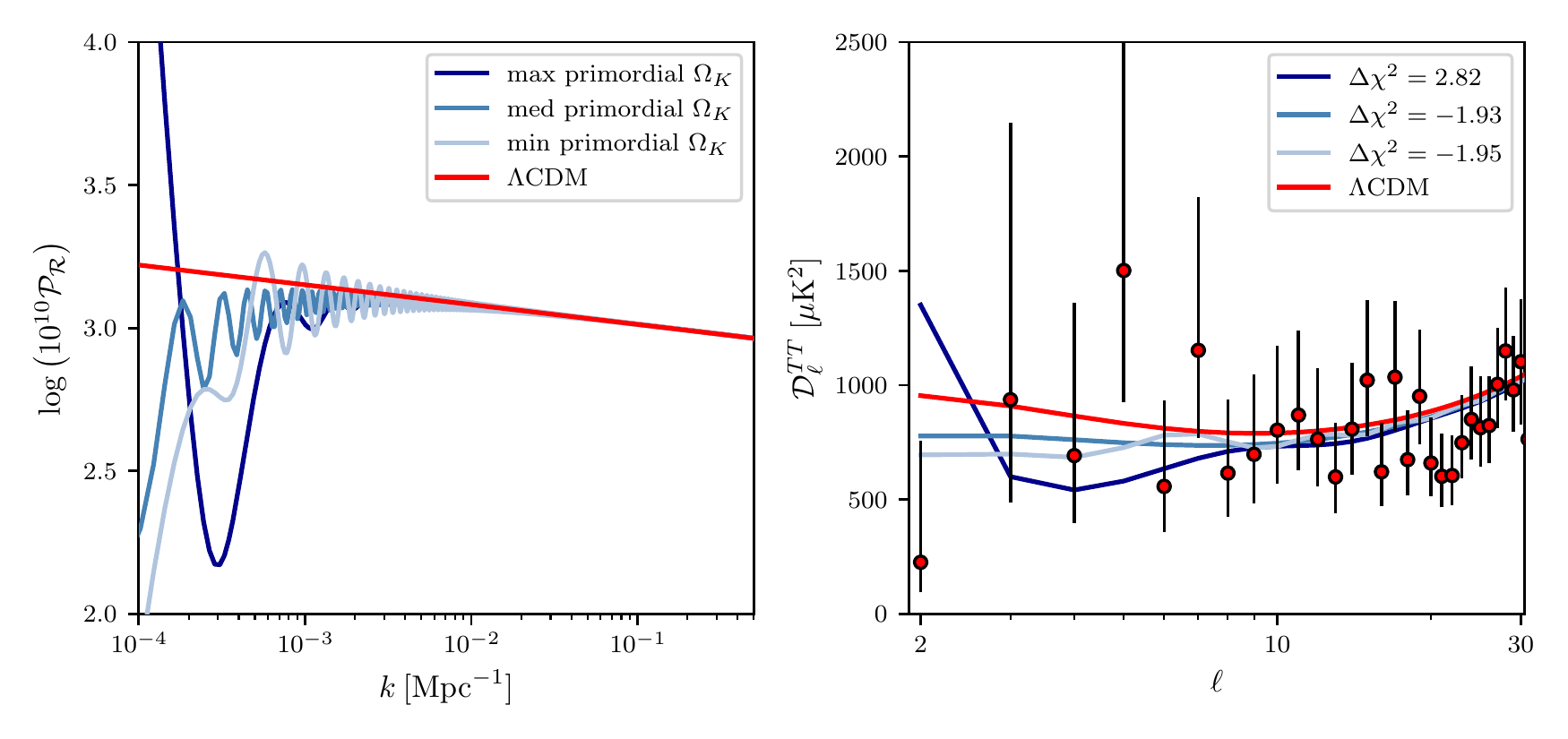}
    \caption{Left: representative best-fit primordial power spectra corresponding to a range of allowed primordial curvatures. \\
    Right: corresponding low-$\ell$ effects on the CMB power spectrum.
}
    \label{fig:powerspec}
\end{figure*} 

\section{\label{sec:powerspectrum}Power Spectrum and Results} 
To ascertain whether the addition of spatial curvature provides a better  description of observations, we aim to compute the power for $\R$. We can write $\R$, $\R'$ in terms of $\zeta$ and $\zeta'$ as
\begin{gather}
    \mathcal{R} = \frac{\zeta}{g} + \frac{K}{\Hu}\frac{f}{g^2c_s^2\mathcal{D}^2}\zeta'\label{eq:RandZeta}\\
    \R'=\frac{1}{g\D^2}\left(\frac{K^2f}{gc_s^2\Hu^2}+(\D^2-K\E)\right)\zeta'+\frac{K}{g\Hu}\zeta \label{eq:RprimeandZeta} 
\end{gather}
for $\mathcal{E}$ as in \cref{eqn:definitions}. Now note that $\zeta$ still describes a family of functions, as $\zeta$ is formulated in terms of a family of functions $f$ and $g$, which are defined by the second-order differential \cref{eq:bina}. For this section, rather than considering $\zeta$ as a physically meaningful perturbation variable in its own right, we shall use it as a means of setting well-motivated initial conditions for $\R$: if one takes $f(\eta_0) = 0$ and $g(\eta_0) = 1$, then $\R = \zeta$ at $\eta_0$. Thus it is appropriate to define initial conditions from $\R$ from the minimised-RSET initial conditions for one such choice in the $\zeta$ family, defined at $\eta=\eta_0$ by $f(\eta_0) = 0$ and $g(\eta_0) = 1$. \Cref{eq:zetarst-conformal,eq:RandZeta,eq:RprimeandZeta} together yield 
\begin{align}
    &\frac{\R'}{\R}\bigg\rvert_{\eta_0} \hspace{-10pt} = \left(1+\frac{Kz^2}{2a^2\kappa_{\mathcal{D}}^2}\right)\left(-i\kappa_{\mathcal{D}} + \frac{a'}{a} - \frac{z'}{z} - \frac{c_s'}{2c_s}- \frac{K}{\Hu}\right) +\frac{K}{\Hu}\bigg\rvert_{\eta_0} \nonumber\\
    &|\R(\eta_0)|^2 = \frac{1}{2c_s\kappa_{\mathcal{D}}z^2}\bigg\rvert_{\eta_0}, \label{eq:Rinitialcond} 
\end{align} 
written in terms of the usual mass variable $z = gz_g$. 

Of note is the independence of \cref{eq:Rinitialcond} on $b$, since it was only introduced as a mathematical tool to construct the desired equation of motion. Furthermore, this result is consistent with the expected physical behaviour of coinciding with the flat case and other initial conditions \cite{kiefer2022power,PhysRevD.78.043534} in the limit of $\kappa_{\mathcal{D}}\to\infty$.

We highlight that this technique would generalise easily, as any perturbation scalar can be written in terms of $\Phi$ and $\Phi'$, and thus equivalently, in terms of $\zeta$, $\zeta'$. Thus for an appropriate choice of initial $f,g$ values, we can set initial conditions for any perturbation scalar by means of minimised RSET on $\zeta$.   

Using an oscillatory solver~\cite{AgocsOscode} and approximating slow roll with $V(\phi) \propto \phi^{4/3}$, one can numerically evolve $\R$ from $t_0$ to the time of mode reentry using \cref{eq:Rinitialcond} and equation of motion (\ref{eqn:mukhanov_sasaki}). For the standard K$\Lambda$CDM universe, we take a parametric form for the primordial power spectrum given by
\begin{equation}
    \mathcal{P}_{\mathcal{R}}^{K\Lambda\text{CDM}}(k) = A_s \left(\frac{k}{k_*} \right)^{n_s-1}.
\end{equation}

The resulting power spectrum for $\R$ using minimised-RSET initial conditions is given in \cref{fig:powerspec}.

\begin{figure*}
    \includegraphics{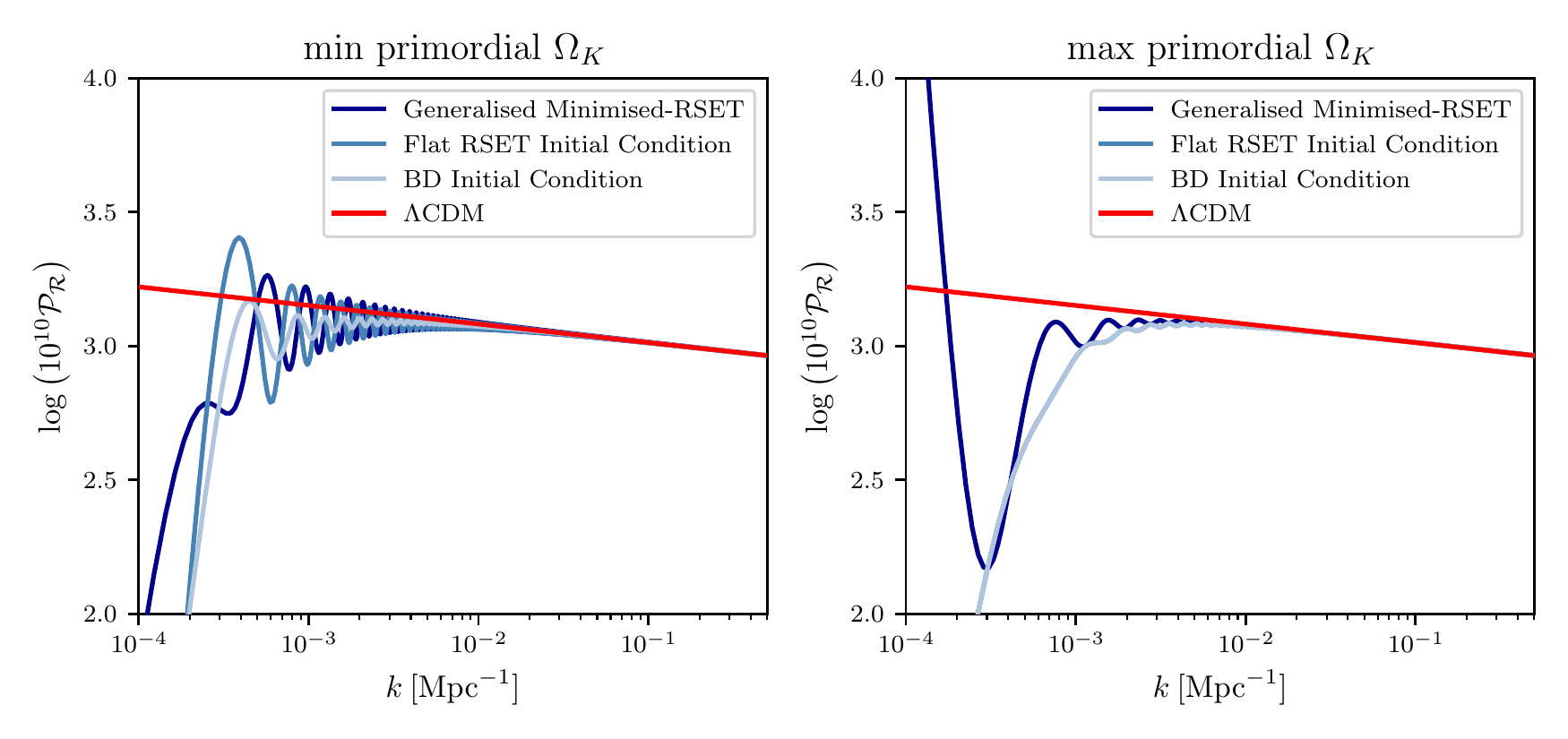}
    \caption{Comparison of the effect of initial conditions on the power spectrum of $\R$. Initial conditions considered are the novel initial conditions \cref{eq:Rinitialcond} for $\R$ by means of generalised-RSET on $\zeta$; a naive application of flat-space RSET conditions computed in Ref.~\cite{Handley_2016}; and Bunch-Davies ICs. The power spectrum is computed for minimum and maximum primordial curvature cases. Note that in the maximum case (right hand graph) the BD condition solution numerically coincides with the naive-flat condition.}
    \label{fig:vacuumcompare}
\end{figure*}

We can compare this with a naive application of the flat-case minimised-RSET conditions, as given in Refs.~\cite{Handley_2016,Handley_2019,Hergt_2022}, as well as with the Bunch-Davies vacuum conditions~\cite{tasi,Handley_2019}. This is shown in \cref{fig:vacuumcompare}, where we see a clear increase in power spectrum oscillations when using the new minimised-RSET conditions derived in this paper, thus hopefully corresponding to better detectability. We also highlight that the differences between vacuum choices is particularly pronounced for larger values of primordial curvature and for lower $k$ modes (where the presence curvature is more relevant), as expected. 

\section{Conclusion}\label{sec:conclusion}
We began by introducing the relevant setup, highlighting work by Ref.~\cite{Handley_2019} in deriving the $K\neq 0$ Mukhanov-Sasaki equation of motion for the gauge-invariant comoving curvature perturbation $\R$. We discussed why this equation of motion prevents setting initial conditions for $\R$ in a well-defined and physically motivated way, and motivated the method of minimised RSET. 

We then propose a novel perturbation scalar $\zeta$, inspired by the case of perfect-fluid filled universe as in Ref.~\cite{brechetfluid}, and with particular emphasis on \cref{eq:zeta0primescalar}. Under the Bardeen equation of motion \cref{eq:bardeenscalar2cs}, describing the evolution of perturbation scalars in our inflationary universe, the equation of motion for $\zeta$ collapses to the desired canonically-normalisable wave equation \cref{eq:finallywaveeq} (i.e., of the same form as the typically considered equation of motion \cref{eq:k0r} for $\R$ when $K=0$). We believe this variable will be an important catalyst for further work on the subject of inflation in curved spacetimes, as its SHO form allows for connection to standard inflationary and quantum field theory literature. 

Building on calculations in Ref.~\cite{Agocs_2020}, we generalise the minimised-RSET vacuum-setting procedure proposed in Ref.~\cite{Handley_2016} to curved spacetimes, allowing us to quantise $\zeta$ (and any variable with analogous equation of motion). Finally, we demonstrate how to use the family of $\zeta$ variables to construct a well-motivated set of initial conditions for $\R$, allowing us to plot the resulting power spectrum. We see changes in $\triangle \chi^2$ in all cases compared to previously studied BD and flat-RSET conditions, particularly noting an increase in power spectrum oscillatory behaviour. 

This work, however, leaves several questions open. First, what is the correct choice of variable for quantization? Ideally there would be a unique, well-motivated theoretical choice, though of course given its impact on the primordial power spectrum, the single degree of freedom in choice of $\zeta$ could also be fit for. Second, the link between curvature and nonadiabaticity given by \cref{eq:entropicpphi,eq:bardeenscalar-pen,eq:zeta0primescalar-pen} should be further analytically explored in the directions of explaining the nonlocality of action \cref{eq:actionR} and identifying the degree of uniqueness of the variable $\zeta$. Finally, it remains to be seen to what extent (if at all) these vacuum states can be constrained using modern cosmological data.


\vfill

\pagebreak

\section*{Acknowledgements}
M.I.~L. thanks the Royal Society and the Cambridge Department of Applied Mathematics and Theoretical Physics (DAMTP) Cambridge Mathematics Placements Programme (CMP) for providing summer bursary funding for this work.  Z.~S. was funded through a summer project by DAMTP CMP. F.J.~A. was supported by STFC and the Flatiron Institute, a division of the Simons Foundation. W.J.~H. was supported by a Royal Society University Research Fellowship.

\newpage
\bibliography{apssamp}

\newpage
\appendix
\section{\label{sec:rstcalc}General minimised RSET calculations for wave-like fields} In this appendix, we aim to relate the equations of motion for $\zeta$ to a free scalar field $\phi$ in the same background, so that we can set initial conditions for $\zeta$ through initial conditions for $\phi$ by RSET. The procedure is invariant to the transformations we use to map $\zeta$ equation of motion (EOM) to $\phi$ EOM, as we will see later in this calculation. Thus, we aim to relate
\begin{equation}
    (z_g\zeta_k)'' + \left(c_s^2(\eta)\kappa_{\mathcal{D}}^2(k) - \frac{z_g(\eta)''}{z_g(\eta)} \right)(z_g\zeta_k) = 0 \label{eq:zeta-eom-kspace}
\end{equation}
to
\begin{equation}
    (a\psi_k)'' + \left(\kappa^2(k) - \frac{a''}{a}\right)(a\psi_k) = 0,
\end{equation}
by means of time redefinition and field rescaling defined in \cref{eq:time-redef-zeta,eq:field-redef-zeta,eq:time-redef-psi,eq:field-redef-psi}.

Time redefinition from $\eta$ to $\eta_{\zeta}$ gives
\begin{align}
    \frac{\d{}}{\d{\eta}} &= \eta_{\zeta}' \frac{\d{}}{\d{\eta_{\zeta}}} \\
    \frac{\d{}^2}{\d{\eta^2}} &= (\eta_{\zeta}')^2\frac{\d{}^2}{\d{\eta_{\zeta}^2}} + \eta_{\zeta}''\frac{\d{}}{\d{\eta_{\zeta}}}
\end{align}
where $'$ denotes differentiation with respect to conformal time $\eta$. Then \cref{eq:zeta-eom-kspace} becomes the equivalent equation of motion for $\chi_{\zeta}$ given by
\begin{align}
    0 &= \partial_{\eta_{\psi}\eta_{\psi}}\chi_{\zeta} + \partial_{\eta_{\zeta}}\chi_{\zeta}\left(\frac{\eta_{\zeta}''}{(\eta_{\zeta}')^2} + 2\frac{\partial_{\eta_{\zeta}}h_{\zeta}}{h_{\zeta}} + 2\frac{\partial_{\eta_{\zeta}}z_g}{z_g} \right) 
    \nonumber \\
    &+ \chi_{\zeta} \left(\frac{\partial_{\eta_{\zeta}}^2h_{\zeta}}{h_{\zeta}} + \frac{\eta_{\zeta}''}{(\eta_{\zeta}')^2}\frac{\partial_{\eta_{\zeta}}h_{\zeta}}{h_{\zeta}} + 2\frac{\partial_{\eta_{\zeta}}h_{\zeta}}{h_{\zeta}}\frac{\partial_{\eta_{\zeta}}z_g}{z_g} + \frac{c_s^2\kappa_{\mathcal{D}}^2}{(\eta_{\zeta}')^2}\right).
\end{align}

We aim to have an equation of motion with no $\partial_{\eta_{\zeta}}\chi_{\zeta}$ terms; i.e., want to choose $h_{\zeta}, \eta_{\zeta}$ such that
\begin{align}
    0 &= \frac{\eta_{\zeta}''}{(\eta_{\zeta}')^2} + 2\frac{\partial_{\eta_{\zeta}}h_{\zeta}}{h_{\zeta}} + 2\frac{\partial_{\eta_{\zeta}}z_g}{z_g} 
    \nonumber \\
    &= \frac{\partial_{\eta_{\zeta}}(C_{\zeta}^2)}{C_{\zeta}^2}
\end{align}
for
\begin{equation}
    C_{\zeta}^2 = \eta_{\zeta}'h_{\zeta}^2z_g^2.
    \label{eq:cw}
\end{equation}
Thus, we fix $C_{\zeta}^2$ to be constant. Then 

\begin{align}
    0 &= \partial_{\eta_{\zeta}\eta_{\zeta}}\chi_{\zeta} 
    \nonumber \\
    &+ \chi_{\zeta}\left( \left(\frac{c_s(\eta_{\zeta})\kappa_{\mathcal{D}}(k)h_{\zeta}^2z_g^2}{C_{\zeta}^2} \right)^2 + \frac{\partial_{\eta_{\zeta}\eta_{\zeta}}h_{\zeta}}{h_{\zeta}} - 2\left( \frac{\partial_{\eta_{\zeta}}h_{\zeta}}{h_{\zeta}}\right)^2 \right) 
\end{align} 
The equivalent formulation for the rescaled and time-redefined $\psi$ field is
\begin{align}
    0 &= \partial_{\eta_{\psi}\eta_{\psi}}\chi_{\psi}
    \nonumber \\
    &+ \chi_{\psi}\left( \left(\frac{\kappa(k)h_{\psi}^2a^2}{C_{\psi}^2} \right)^2 + \frac{\partial_{\eta_{\psi}\eta_{\psi}}h_{\psi}}{h_{\psi}} - 2\left( \frac{\partial_{\eta_{\psi}}h_{\psi}}{h_{\psi}}\right)^2 \right) \label{eq:chipsieom}
\end{align}
for constant
\begin{equation}
    C_{\psi}^2 = \eta_{\psi}'h_{\psi}^2a^2.
    \label{eq:cpsi}
\end{equation} 

As argued in Ref.~\cite{Agocs_2020},
\begin{equation}
    \frac{\partial_{\eta_{\psi}\eta_{\psi}}h_{\psi}}{h_{\psi}} - 2\left( \frac{\partial_{\eta_{\psi}}h_{\psi}}{h_{\psi}}\right)^2 = \frac{\partial_{\eta_{\zeta}\eta_{\zeta}}h_{\zeta}}{h_{\zeta}} - 2\left( \frac{\partial_{\eta_{\zeta}}h_{\zeta}}{h_{\zeta}}\right)^2 \label{eq:masseq1}
\end{equation}
can always be made to be true for suitable choices of $h_{\zeta}, \eta_{\zeta}, h_{\psi}, \eta_{\psi}$. 

Hence the only remaining condition to be fixed is
\begin{equation}
    \frac{\sqrt{c_s}h_{\zeta}z_g}{C_{\zeta}} = \frac{h_{\psi}a}{C_{\psi}},
    \label{eq:masseq2}
\end{equation}
having shifted the wave number for $\psi$ so that
\begin{equation}
    \kappa_{\mathcal{D}}(k_{\text{shift}}) = \kappa(k).
\end{equation} 

Now promote $\chi_{\psi}$ to an operator, as is standard in canonical quantisation; then $\psi$ may be expanded in terms of creation and annihilation operators as
\begin{equation}
    \psi = \int \frac{\d{}^3k}{(2\pi)^3}h_{\psi}(\eta_{\psi})\left( \hat{a}_k\chi_{\psi,k}Y_k(\text{\textbf{x}}) + \hat{a}_k^{\dagger}\chi^*_{\psi,k}Y^*_k(\text{\textbf{x}})\right), \label{eq:curvedpsifruz}
\end{equation}
where $Y_k(x)$ is the eigenfunction of the curved spatial Laplacian, namely
\begin{equation}
    \nabla^2 Y_k(\text{\textbf{x}}) = -\kappa^2(k)Y_k(\text{\textbf{x}}).
\end{equation} 
The renormalised $T_{00}$ component is then given by
\begin{equation}
    \langle 0 | T_{00} | \rangle_{\text{ren}} = \lim_{\text{\textbf{x}}\to \text{\textbf{x}}'} \mathcal{D}_{00} G(\text{\textbf{x}},\text{\textbf{x}}') - \Tilde{T}, 
\end{equation}
where
\begin{equation}
    G(\text{\textbf{x}},\text{\textbf{x}}') = \frac12 \langle 0 | \psi(\text{\textbf{x}})\psi(\text{\textbf{x}}') + \psi(\text{\textbf{x}}')\psi(\text{\textbf{x}}) | 0\rangle.
\end{equation}
Crucially, the de-Witt Schwinger geometrical terms given by $\Tilde{T}$ are independent of the field variables $\chi_{\psi}$ (or $\psi$), shown in \citet{birrell_davies_1982} (Chapter 6, Section 6.4). Thus, we expand
\begin{align}
    2G(\text{\textbf{x}},\text{\textbf{x}}') &= \int \frac{\d{}^3k}{(2\pi)^3} h(\eta_{\psi})h(\eta_{\psi}')\chi(\eta_{\psi})\chi^*(\eta_{\psi}')Y(\text{\textbf{x}})Y^*(\text{\textbf{x}}') \nonumber \\
    &+ h(\eta_{\psi})h(\eta_{\psi}')\chi(\eta_{\psi}')\chi^*(\eta_{\psi})Y(\text{\textbf{x}}')Y^*(\text{\textbf{x}}) )
\end{align}
where for brevity, $h = h_{\psi}$, $\chi = \chi_{\psi,k}$ and $Y = Y_k$. Then
\begin{align}
    \langle 0 | T_{00} | \rangle_{\text{ren}} &= \Tilde{T} + \frac12 \int \frac{\d{}^3k}{(2\pi)^3} \frac{h^2\kappa^2}{(\partial_{\eta}\eta_{\psi})^2}\chi\chi^*YY^* \nonumber \\
    &+ h^2\left(\partial_{\eta_{\psi}}\chi + \frac{\partial_{\eta_{\psi}}h}{h}\chi\right)\left(\partial_{\eta_{\psi}}\chi^* + \frac{\partial_{\eta_{\psi}}h}{h}\chi^*\right)YY^* \label{eq:t00}
\end{align}
The spherical-harmonic-like functions $Y_k(x)$ are normalised as
\begin{equation}
    \int \d{}^3x \sqrt{\text{det}(c_{ij})}Y_k(x)Y_p(x) = \delta(k-p)
\end{equation} 

so that the canonical commutation relation for the redefined field is given by
\begin{equation}
    (\partial_{\eta_{\psi}}\chi_{\psi,k})\chi_{\psi,k}^* - (\partial_{\eta_{\psi}}\chi_{\psi,k}^*)\chi_{\psi,k} = -\frac{i}{C_\psi^2}. \label{eq:canonicalcommutation}
\end{equation}
Finally, minimising \cref{eq:t00} with respect to field variables $\chi_{\psi,k}, \chi_{\psi,k}^*$ and their derivatives $\partial_{\eta_{\psi}}\chi_{\psi,k}, \partial_{\eta_{\psi}}\chi_{\psi,k}^*$ subject to the constraint \cref{eq:canonicalcommutation} gives
\begin{align}
    \partial_{\eta_{\psi}} \ln(\chi_{\psi,k}) &= -\frac{i\kappa(k)}{\eta_{\psi}'} - \partial_{\eta_{\psi}} \ln(h_{\psi}), \\
    |\chi_{\psi,k}^2| &= \frac{1}{2\kappa h_{\psi}^2a^2}.
\end{align}

The equation of motion for $\chi_{\zeta}$ in terms of $\eta_{\zeta}$ at $k_{\text{shift}}(k)$ is now identical to the equation of motion for $\chi_{\psi}$ in terms of $\eta_{\psi}$ at $k$. Thus we have equivalence of solutions
\begin{equation}
    \chi_{\psi}(k) = \chi_{\zeta}(k_{\text{shift}}).
\end{equation}

Substituting $\chi_{\psi}(k) = \chi_{\zeta}(k_{\text{shift}})$, using $h_{\psi}$ for $h_{\zeta}$, and converting to conformal time gives
\begin{equation}
     \frac{\zeta'}{\zeta} = -i\kappa_{\mathcal{D}} + \frac{a'}{a} - \frac{z_g'}{z_g} - \frac12\frac{c_s'}{c_s}.
 \end{equation}
 
Introducing 2 arbitrary time redefinitions and 2 arbitrary field rescalings gives 4 degrees of freedom. Thus far, we still have one remaining degree of freedom: \cref{eq:masseq1,eq:masseq2} each fix one degree of freedom, and \cref{eq:cw,eq:cpsi} \textit{together} fix another degree of freedom (as the \textit{values} of $C_{\zeta},C_{\psi}$ are arbitrary). Thus we are free to set $C_{\zeta} = C_{\psi}$. In conclusion, we can initialise $\zeta$ at $\eta = \eta_0$ by
\begin{align}
    |\zeta(\eta_0)|^2 &= \frac{1}{2c_s(\eta_0)z_g^2(\eta_0)\kappa_{\mathcal{D}}}, 
    \nonumber \\
    \frac{\zeta'}{\zeta}(\eta_0) &= -i\kappa_{\mathcal{D}} + \frac{a'}{a}(\eta_0) - \frac{z_g'}{z_g}(\eta_0) - \frac12\frac{c_s'}{c_s}(\eta_0),
\end{align}
or at $t = t_0$ by
\begin{align}
    |\zeta(t_0)|^2 &= \frac{1}{2c_s(t_0)z_g^2(t_0)\kappa_{\mathcal{D}}}, \nonumber \\
    \frac{\dot{\zeta}}{\zeta}(t_0) &= -i\frac{1}{a(t_0)}\kappa_{\mathcal{D}} + \frac{\dot{a}}{a}(t_0) - \frac{\dot{z_g}}{z_g}(t_0) - \frac12\frac{\dot{c_s}}{c_s}(t_0).
\end{align} The forms of these equations do not depend on the canonical transformations used to relate $\zeta$ to $\phi$, i.e., $\eta_{\zeta},\eta_{\phi},h_{\zeta},h_{\phi},$ which proves that initial conditions set by RSET are invariant under such transformations even in the $K\neq 0$ case. 

\section{\label{sec:variable_explained}Motivation and construction of $\zeta$} 
The starting point of this section is \cref{eq:bardeenscalar2cs}. For the rest of this section we will denote with capital letters functions of $\eta$, and indicate $x$ (or $k$) dependence by subscripts. With this in mind, we write \cref{eq:bardeenscalar2cs} as  
\begin{equation}
    \Phi''+A_1\Phi'+ B_1\Phi - D_{k}\Phi = 0.
    \label{eq:bardeenappendix}
\end{equation}

Then, we would like to construct a variable 
\begin{equation}
    \zeta = A\Phi+B\Phi',
\end{equation}
for which we will require the following:
\begin{enumerate}
    \item $A$ and $B$ are not to have any $x$ (or $k$) dependence.
    \item The resulting equation for zeta is to be of the form
    \begin{equation}
    \zeta''+M\zeta'+D_{2;k}\zeta=0,    
    \label{eq:zetaappendixeq}
    \end{equation}
    where, for the time being, we do not require $D_{2;k}$ to be the same as $D_k$.
\end{enumerate}
Based on these requirements we can write 
\begin{align}
    \zeta&=A\Phi+B\Phi',\\
    \zeta'&=C\Phi+P\Phi'-BD_k\Phi,\\
    \zeta''&=N\Phi+L\Phi'-PD_k\Phi-B'D_k\Phi-BD_k'\Phi-BD_k\Phi',
\end{align}
where we have defined
\begin{align}
    C &= A'-BB_1,\\
    \label{eq:PPeqn}P &= A+B'-BA_1,\\
    N &= C'-PB_1,\\
    L &= C+P'-PA_1.
\end{align}
Substituting this into \cref{eq:zetaappendixeq},  then matching coefficients of $\Phi,\Phi'$, and separately the coefficients of the $k$-dependent components of $\Phi,\Phi'$, we have
\begin{align}
    N &= -MC\label{eq:leftover1}\\
    L &= -MP\label{eq:leftover2}\\
    BD_{2;k}&=BD_k\\
    \label{eq:appmaineq}0&= -PD-B'D-BD'+AD_{2;k}-MBD.
\end{align}
Therefore, we must have that $D_{2;k}$ is the same as $D_k$. However, we had a degree of freedom in choosing $D_k$ when writing down \cref{eq:bardeenappendix}. But from \cref{eq:appmaineq} we have
\begin{equation}
    A - MB - P - B' = B\frac{D_k'}{D_k},
\end{equation}
meaning that we must have $D_k$ being a product of a function of $\eta$ by a function of $k$. E.g. when writing down \cref{eq:bardeenappendix}, we could pick $D_k = c_s^2\D^2$ or $c_s^2\nabla^2$, but not $D_k=c_s^2\nabla^2+K$.
The above asserts $D_{2;k}$ to be the same as $D_k$; therefore, all that is needed to find the form of the wave equation is identifying $M$ while satisfying \cref{eq:leftover1,eq:leftover2,eq:appmaineq}. Let $M=\frac{2(z\delta)'}{z\delta}$, i.e., a mass rescaling for our new variable. Then, from \cref{eq:appmaineq,eq:PPeqn}, we have
\begin{equation}
    \frac{2B'}{B}=-M+A_1-\frac{D_k'}{D_k}
\end{equation}
If we pick \cref{eq:zetadef} as the general form for our equation, i.e., swap from $A$ and $B$ to $g$ and $f$, we must have that $\delta = \frac{1}{g}$. We can further simplify \cref{eq:leftover2,eq:leftover1} to be the following two equations that are linear in $C$ and $P$:
\begin{align}
    \left(\frac{Cz^2}{g^2}\right)'&=-\frac{Pz^2}{g^2}\frac{(\rho+p)'}{\rho+p},\label{eq:triv2}\\
    \left(\frac{Pz^2(\rho+p)}{g^2}\right)'&=\frac{Cz^2(\rho+p)}{g^2},\label{eq:triv1}
\end{align}
where for the specific choice of $D_k = c_s^2\D^2$
\begin{align}
    P &= \frac{K}{\Hu}-\frac{f}{g}\frac{K}{\Hu}+\frac{g'}{g}\\
    C &=K+\frac{g'}{g}\left(\Hu+\frac{a^2(\rho+p)}{2\Hu}\right)\\&\quad+\frac{K}{\Hu}\frac{f}{g}\left(2\Hu+\frac{(\rho+p)'}{(\rho+p)}-\frac{f'}{f}\right).\nonumber
\end{align}
Solving these equations in general is rather difficult, since they are two coupled, nonlinear, second order differential equations for $f$ and $g$. We will aim to simplify our our calculations by making an ansatz. If we are to pick $f$ and $g$, such that $P=C=0$, then equations \cref{eq:triv1,eq:triv2} are trivially satisfied. This gives us exactly two restrictions on our functions $f$ and $g$:
\begin{align}
    f&=g'\frac{\Hu}{K}+g\label{eq:defining1}\\
    \frac{f'}{f} &= K+\frac{g'}{g}\left(\Hu+\frac{a^2(\rho+p)}{2\Hu}\right)+\frac{K}{\Hu}\frac{f}{g}\left(2\Hu+\frac{(\rho+p)'}{(\rho+p)}\right)\label{eq:defining2}
\end{align}
Thus, if we can find solutions to \cref{eq:defining1,eq:defining2}, then equation collapses to the required wave equation \cref{eq:finallywaveeq}. Setting 
\begin{equation}
    g(\eta)=\frac{a}{\Hu}\frac{b'}{b^2}    
\end{equation}
reduces \cref{eq:defining2} to 
\begin{equation}
b\left(\frac{1}{b}\right)''=a\left(\frac{1}{a}\right)''+K,
\end{equation}
which gives a one-parameter family of solutions for $\zeta$, since one of the free parameters in this second order differential equation is due to arbitrary scaling of $b$ (or equivalently $g$). We can see from this that this ansatz is sufficient for a SHO equation for $\zeta$, but it remains an open question whether it is necessary i.e., whether $P=C=0$ are the only solutions to \cref{eq:triv1,eq:triv2}.
\newpage
\end{document}